\documentclass{ieeeaccess}
\usepackage{cite}
\usepackage{amsmath,amssymb,amsfonts}
\usepackage{algorithmic}
\usepackage{graphicx}
\usepackage{textcomp}
\usepackage{url}        
\usepackage{booktabs}      
\usepackage{amsfonts}       
\usepackage{nicefrac}      
\usepackage{microtype}  
\usepackage{xcolor}   
\usepackage{graphicx} 
\usepackage{subcaption} 
\usepackage{multirow}
\usepackage{hyperref}
\usepackage{booktabs}
\usepackage{tabularx}
\usepackage{multicol}
\usepackage{mdframed}
\usepackage{bm}
\usepackage{pict2e}

\newsavebox{\ORCIDlogo}
\savebox{\ORCIDlogo}{%
\setlength{\unitlength}{\dimexpr 1em/256\relax}%
\begin{picture}(256,256)%
  \color[HTML]{A6CE39}\put(128,128){\circle*{256}}%
  \color{white}%
  \put(78.6,199.2){\circle*{20}}%
  \moveto(70.9,176,9)\lineto(86.3,176,9)\lineto(86.3,69.8)\lineto(70.9,69.8)%
  \closepath\fillpath%
  \moveto(108.9,176.9)\lineto(150.5,176.9)%
  \curveto(190.1,176.9)(207.5,148.6)(207.5 ,123.3)%
  \curveto(207.5,95,8)(186,69.7)(150.7,69.7)%
  \lineto(108.9,69.7)%
  \closepath\fillpath%
  \color[HTML]{A6CE39}%
  \moveto(124.3,83.6)\lineto(148.8,83.6)%
  \curveto(183.7,83.6)(191.7,110.1)(191.7,123.3)%
  \curveto(191.7,144.8)(178,163)(148,163)%
  \lineto(124.3,163)%
  \closepath\fillpath%
\end{picture}%
}

\newcommand\orcidicon[1]{\href{https://orcid.org/#1}{\usebox{\ORCIDlogo}}}

\usepackage{hyperref}

\newmdenv[
  backgroundcolor=gray!10,
  linecolor=black!75,
  linewidth=0.5pt,
  roundcorner=6pt,
  topline=true,
  frametitlebackgroundcolor=black!75,
  frametitlerule=true,
  frametitlefont=\bfseries\color{white},
]{mybox}

\newmdenv[
  backgroundcolor=gray!5,
  linecolor=gray!75,
  linewidth=0.3pt,
  roundcorner=6pt,
]{promptbox}

\makeatletter
\AtBeginDocument{\DeclareMathVersion{bold}
\SetSymbolFont{operators}{bold}{T1}{times}{b}{n}
\SetSymbolFont{NewLetters}{bold}{T1}{times}{b}{it}
\SetMathAlphabet{\mathrm}{bold}{T1}{times}{b}{n}
\SetMathAlphabet{\mathit}{bold}{T1}{times}{b}{it}
\SetMathAlphabet{\mathbf}{bold}{T1}{times}{b}{n}
\SetMathAlphabet{\mathtt}{bold}{OT1}{pcr}{b}{n}
\SetSymbolFont{symbols}{bold}{OMS}{cmsy}{b}{n}
\renewcommand\boldmath{\@nomath\boldmath\mathversion{bold}}}
\makeatother

\def\BibTeX{{\rm B\kern-.05em{\sc i\kern-.025em b}\kern-.08em
    T\kern-.1667em\lower.7ex\hbox{E}\kern-.125emX}}

\begin{document}
\history{Date of publication 2 December 2025,
date of current version 22 December 2025.}
\doi{10.1109/ACCESS.2025.3638251}

\title{ProveRAG: Provenance-Driven Vulnerability Analysis with Automated Retrieval-Augmented LLMs}
\author{\uppercase{Reza Fayyazi}\authorrefmark{1}  \orcidicon{0009-0001-5323-9504} \IEEEmembership{Graduate Student Member, IEEE},
\uppercase{Stella Hoyos Trueba}\authorrefmark{2},  \uppercase{Michael Zuzak}\authorrefmark{1} \IEEEmembership{Member, IEEE}, \uppercase{Shanchieh Jay Yang}\authorrefmark{3} \IEEEmembership{Senior Member, IEEE}}

\address[1]{Department of Computer Engineering, Rochester Institute of Technology, Rochester, NY 14623, USA}
\address[2]{Department of Computer Science, Karlsruhe Institute of Technology, Karlsruhe 76131, Germany}

\address[3]{Informatics and Applied Technology Institute, Gonzaga University, Spokane, WA 99258, USA}
\tfootnote{This material is based upon work supported by the National Science Foundation under Grant Nos. 2344237, 2228001, and DGE-2125362.}

\markboth
{R. Fayyazi et al. \headeretal: ProveRAG: Provenance-Driven Vulnerability Analysis With Retrieval-Augmented LLMs}
{R. Fayyazi et al. \headeretal: ProveRAG: Provenance-Driven Vulnerability Analysis With Retrieval-Augmented LLMs}

\corresp{Corresponding author: Reza Fayyazi (e-mail: rf1679@rit.edu).}

\begin{abstract}
In cybersecurity, security analysts constantly face the challenge of mitigating newly discovered vulnerabilities in real-time, with over 300,000 vulnerabilities identified since 1999. The sheer volume of known vulnerabilities complicates the detection of patterns for unknown threats. While LLMs can assist, they often hallucinate and lack alignment with recent threats. Over 40,000 vulnerabilities have been identified in 2024 alone, which are introduced after most popular LLMs' (e.g., GPT-5) training data cutoff. This raises a major challenge of leveraging LLMs in cybersecurity, where accuracy and up-to-date information are paramount. Therefore, we aim to improve the adaptation of LLMs in vulnerability analysis by mimicking how an analyst performs such tasks. We propose \textbf{ProveRAG}, an LLM-powered system designed to assist in rapidly analyzing vulnerabilities with automated retrieval augmentation of web data while self-evaluating its responses with verifiable evidence. ProveRAG incorporates a self-critique mechanism to help alleviate the omission and hallucination common in the output of LLMs applied in cybersecurity applications. The system cross-references data from verifiable sources (NVD and CWE), giving analysts confidence in the actionable insights provided. Our results indicate that ProveRAG excels in delivering verifiable evidence to the user with over 99\% and 97\% accuracy in exploitation and mitigation strategies, respectively. 
ProveRAG guides analysts to secure their systems more effectively by overcoming temporal and context-window limitations while also documenting the process for future audits.
\end{abstract}


\titlepgskip=-21pt

\maketitle

\section{Introduction}
In today's fast-paced cybersecurity landscape, security analysts are continuously challenged to understand and mitigate newly discovered threats in real-time \cite{okutan2023empirical}. Common Vulnerabilities and Exposures (CVEs) \cite{CVE} serve as a critical framework for identifying and detecting known vulnerabilities. With over 300,000 CVEs cataloged since 1999, the sheer volume makes it increasingly difficult and time-consuming for analysts to detect patterns and respond effectively to emerging threats. Public vulnerability databases, such as the NIST National Vulnerability Database (NVD) \cite{nvd2024}, predominantly rely on labor-intensive manual processes to analyze and categorize vulnerabilities. This reliance on human effort can cause delays in threat identification and response, which makes it difficult to keep pace with the rapidly evolving threat landscape \cite{okutan2023empirical}. 

For example, the \textit{CVE-2024-0302} is a critical vulnerability that, if left unchecked, could lead to deserialization of untrusted data. To mitigate this vulnerability, fields should be made transient to be protected from deserialization. 
However, protecting an enterprise network necessitates discovering which vulnerabilities are relevant by combing through over 300,000 threats aggregated in these databases, with more added each day. Separate mitigation databases, such as the Common Weakness Enumeration (CWE) \cite{cwe2024}, must also be explored to identify, implement, and deploy mitigations to the system. Due to the sheer volume of information, manually identifying and mitigating threats is challenging, and automating the identification and mitigation process is essential to reduce risk and improve overall resilience.  

Large Language Models (LLMs) have been widely adopted in cybersecurity applications \cite{mitra2024localintel, ullah2024llms, cheshkov2023evaluation, deng2024pentestgpt}. They have been particularly effective in assisting with the analysis of Cyber Threat Intelligence \cite{mitra2024localintel, perrina2023agir}, supporting LLM-assisted attacks  \cite{sharma2023impact, gupta2023chatgpt, deng2024pentestgpt}, and improving vulnerability detection \cite{cheshkov2023evaluation, ullah2024llms}. 

However, prior work has identified three aspects that limit the broader adoption of LLMs to cybersecurity applications \cite{huang2023survey, rawte2023troubling, zhu2024evaluating, liu2024lost}, which we aim to address in this work. These limitations are: 1) \textit{Hallucination:} the tendency to generate misleading information that appears plausible but is factually incorrect. Some works have shown the consequences of this problem \cite{huang2023survey, rawte2023troubling}. 2) \textit{Temporal Knowledge:} the failure of LLMs to respond accurately when they are exposed to data that come after the cutoff point of their training data \cite{zhu2024llms, zhu2024evaluating}. In cybersecurity, daily emerging vulnerabilities—over 40,000 identified in 2024 alone—pose serious risks, especially if they appear after LLMs' training data cutoff. Retrieval Augmented Generation (RAG) techniques have been proposed to address this problem \cite{borgeaud2022improving} by retrieving relevant external information. However, adoption of RAG often leads to  3) \textit{Limited Context Windows:} which restrict the number of tokens that can be provided to prompt the model, thus limiting RAG-based temporal alignment approaches \cite{liu2024lost}. 
These limitations constitute a major challenge of leveraging LLMs in cybersecurity, where accuracy, time, and up-to-date information is essential. 

In this work, we aim to improve the adaptation of LLMs in vulnerability analysis. To do so, we propose \textbf{ProveRAG}, an advanced LLM-powered system designed to assist security teams in rapidly analyzing CVEs with verifiable evidence. 
What sets ProveRAG apart is its ability to cross-reference, rationalize, and validate information from authoritative sources, such as NVD and CWE. By doing so, we demonstrate that ProveRAG exhibits fewer instances of omission and hallucination through self-critiquing its own response.

A key innovation in ProveRAG is its incorporation of \textbf{provenance} into its analysis. By documenting the source of every piece of information, ProveRAG \textbf{reveals} the hallucinations and inaccuracies common in LLMs' responses and gives security analysts the confidence to make informed decisions. 
For example, when suggesting mitigation strategies for vulnerabilities like \textit{CVE-2024-0302}, ProveRAG reveals pieces of data from trusted databases (such as NVD and CWE) that led to its generated response. This provides a clear audit trail that can be reviewed and verified, thus fostering trust and enabling effective action.

We test ProveRAG on 2024 CVEs, and we compare with two different baseline approaches: 1) prompt-only and 2) chunking retrieval during generation We compare these baselines to our proposed summarizing technique (integrated into ProveRAG). The prompt-only technique is directly asking the LLM details about a specific CVE exploitation and mitigation strategies with data after its training cutoff knowledge. For the chunking technique, we enhance information retrieval by splitting and embedding the retrieved content into smaller chunks. However, this approach requires a very large context window and can include redundant information that might mislead the LLM.

Therefore, we design a summarizing technique to process CVEs, inspired by \cite{kim2024sure, jiang2024tc} in an automatic way by mimicking how security analysts perform vulnerability analysis. To do so, we start by exploring the NVD website and summarize the content with respect to a specific CVE-ID. Next, we move to the associated CWE link, and further to the relevant references (hyperlinks) to provide summaries with respect to the exploitation and mitigation questions (emulating the typical workflow of a security analyst). A key advantage of the summarizing technique is its efficiency in handling large-scale data without relying extensively on document chunking, which enhances both retrieval quality and the accuracy of responses (while handling the context window limitation). This makes the system particularly suited to better guide analysts in a timely manner while addressing more complex cybersecurity queries - such as those related to CVE mitigation that require integrating more external information. 

ProveRAG has the ability to provide verifiable evidence (TPs), while also \textbf{revealing} hallucinations (FPs) and failed retrievals (FNs) to determine whether a generated response is accurate and trustworthy (the provenance part of the proposed framework). Furthermore, we demonstrate that by providing the LLM with highly-structured context (Aqua Vulnerability Database \cite{aqua}), ProveRAG's performance increases significantly. Our testing shows that ProveRAG achieves outstanding detection accuracy, with 99\% for exploitation and 97\% for mitigation strategies. 

ProveRAG not only enables more reliable decision-making but also documents its processes to ensure a robust audit trail for future reviews. For evaluation, we ensure comprehensive coverage of all relevant information for each CVE by incorporating external resources beyond the NVD and CWE. Finally, we show that the proposed ProveRAG framework generalizes across different models by comparing \textit{OpenAI's gpt-4o-mini} and \textit{Meta's Llama-3.1-8B} \cite{llama31}. This allows security analysts to rapidly navigate the massive amounts of data to identify relevant cybersecurity vulnerabilities applicable to their enterprise system and develop appropriate mitigation strategies.
The contributions of our work are as follows:

\begin{itemize}

\item We propose ProveRAG, an LLM-powered system developed to self-critique its generated responses with rationale, by providing pieces of evidence from verifiable sources that led to its generation (i.e., provenance). 

\item We apply ProveRAG to perform vulnerability analysis, a critical application that requires retrieval and verification of new, unseen knowledge.

\item We demonstrate that ProveRAG not only can \textbf{show proof} for its responses (TPs), but also \textbf{reveals} hallucinations (FPs) and failed retrievals (FNs) with evidence. 

\item We show that ProveRAG effectively respond to unseen data while relaxing the limitations imposed by context windows, leading to high \#TPs (99\% accuracy) and low \#FPs and \#FNs, even when answering complex queries (e.g., vulnerability mitigation).

\item We evaluate ProveRAG on two LLMs, \textit{gpt-4o-mini} and \textit{Llama-3.1-8B}, to support the generalizability of the proposed framework.

\end{itemize}

\section{Background \& Related Work}

\subsection{Large Language Models}
The significant advancements in Large Language Models is largely attributed to the integration of the Transformer architecture \cite{vaswani2017attention}. These models are characterized by their extensive training on vast and diverse datasets, leading them to generate coherent and contextually relevant text that closely mirrors human language. OpenAI's GPT-4o \cite{OpenAI} and Meta's LLama-3 \cite{dubey2024llama} are leading examples in this category. The ability of these models to manage diverse and complex tasks has been extensively documented in the literature \cite{min2023recent, Zhao2023}. However, despite the efficacy of LLMs in diverse tasks, these models have shown to be prone to hallucination in their responses \cite{huang2023survey, rawte2023troubling}.

\subsection{Retrieval Augmented Generation}
The hallucination issue of LLMs stems from their reliance on pre-trained knowledge, which causes another limitation and that is the temporal scope of their training data. Such limitations pose significant challenges in fields requiring high precision, such as cybersecurity. To address these challenges and provide more factual knowledge, RAG was introduced \cite{borgeaud2022improving}. RAG effectively combines search algorithms with LLM prompting: the model is prompted to answer a query using relevant information retrieved from external resources as context. This enhances LLMs' ability to generate accurate and contextually relevant responses.

Several advanced RAG techniques have been proposed to further refine this integration \cite{jiang2023flare, su2024dragin, jeong2024adaptive,selfrag,kim2024sure, jiang2024tc}. FLARE \cite{jiang2023flare} improves model accuracy by predicting the next sentence and using low-confidence tokens as queries to re-retrieve relevant documents.
DRAGIN \cite{su2024dragin} leverages the LLM's real-time information needs with self-attention to decide what and when to trigger retrieval. Adaptive-RAG \cite{jeong2024adaptive} employs a smaller LLM as a classifier to assess query complexity and subsequently selects the most appropriate retrieval strategy. 
Self-RAG \cite{selfrag} employs self-reflection for retrieved documents to improve generation quality. However, Self-RAG trains two models--a critic model and a generator model-- to assess its predictions. Our approach avoids any fine-tuning of LLMs, ensuring an automatic process designed to match the rapid pace in cybersecurity applications. Summarizing Retrievals (SuRe) \cite{kim2024sure} constructs summaries of retrieved passages for multiple answer candidates and confirms the most plausible answer by evaluating and ranking the generated summaries. TC-RAG \cite{jiang2024tc} integrates a memory stack system with backtracking and summarization functions to correct errors and eliminate redundant information.  Despite these advancements, all these methods face the challenge of effectively showcasing the evidence underlying their generated responses.

\subsection{LLMs in Vulnerability Analysis}

There are multiple works that integrated the use of LLMs in vulnerability analysis \cite{khare2023understanding, cheshkov2023evaluation, du2024vul, lu2024grace}. 
Anton C. et al. \cite{cheshkov2023evaluation} used GPT models to identify CWE vulnerabilities in Java code and revealed that LLMs struggle with detecting vulnerable code. In contrast, Avishree K. et al. \cite{khare2023understanding} demonstrated with multiple prompt strategies that LLMs can outperform deep learning models in vulnerability detection when carefully prompted, which indicates the importance of prompt engineering. Building on this, Lu et al. \cite{lu2024grace} introduced GRACE, a framework using in-context learning and graph-structured information with LLMs, which showed improvements in vulnerability detection on C/C++ datasets. 
Du et al. \cite{du2024vul} introduced Vul-RAG,  LLM-based vulnerability detection technique that constructs a comprehensive knowledge base by extracting multi-dimensional insights from existing CVE instances. However, the reliance on human analysis in their approach makes it a manual time-consuming work and limits its efficiency in real-time vulnerability detection. In addition, PentestGPT \cite{deng2024pentestgpt} has shown capabilities in assisting in penetration testing. While it does not address the same problem this paper focuses on, the authors noted that LLMs are constrained by limited token sizes (i.e., context window), preventing long-term memory retention. This work builds on this observation and develop efficient RAG to bring LLM's attention to relevant contents.

\section{Motivation \& Problem Formulation}

\begin{figure*}[t]
\centering
\includegraphics[scale=0.41]{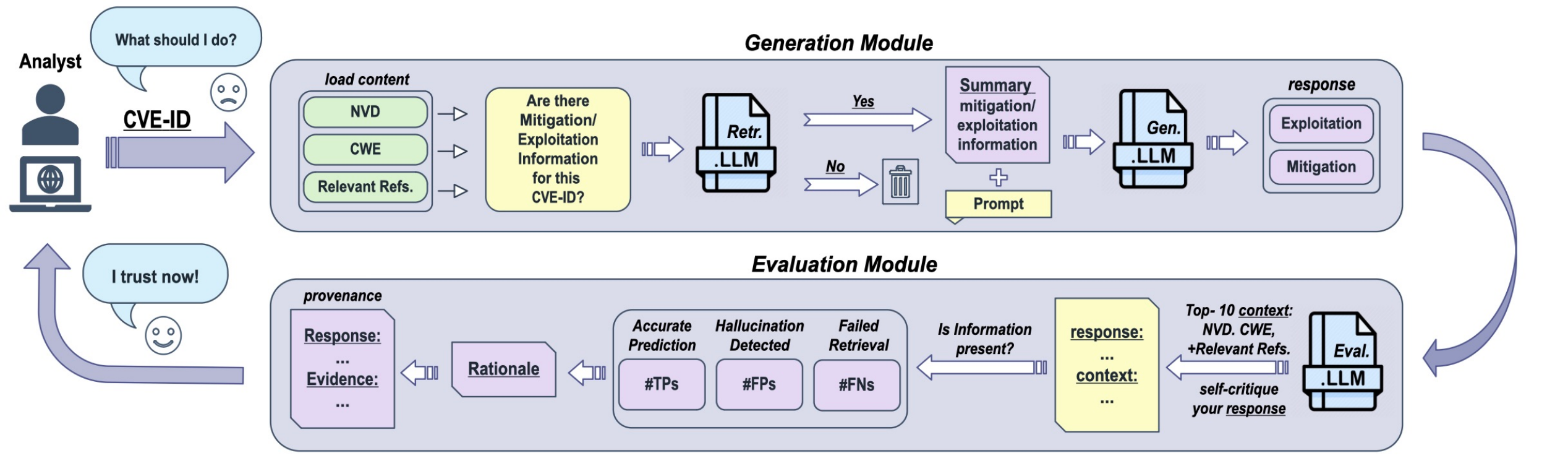}
\caption{ProveRAG: automated retrieval-augmented generation \& self-critique provenance.}
\label{fig:proverag_architecture}
\end{figure*}

Vulnerability analysis involves evaluating a system's configuration to identify potential security risks. By assessing attack surfaces and risk factors, it helps uncover weak points in the system that require effective mitigation to prevent exploitation.
Traditional approaches to vulnerability analysis typically involve manual assessment and the use of static databases, which can be both time-consuming and prone to human error. Moreover, while existing systems can assess vulnerabilities, they often fall short in providing detailed, actionable insights necessary for effective mitigation. This gap in the process leaves security teams with a critical need for more advanced systems capable of assessing threats, generating comprehensive mitigation scenarios, and documenting evidences of their recommendations. The motivation behind this work stems from these challenges. We seek to address these limitations by answering the following research questions:

\noindent \textbf{RQ1: How can the knowledge cutoff of LLMs be addressed in the context of continuously emerging cybersecurity threats?}
One of the key problems we address is the knowledge cutoff (temporal limitation) of LLMs. With the constant emergence of new vulnerabilities, it is crucial that security systems remain effective even when dealing with data that falls outside the LLM's training window. Our system, ProveRAG, is designed to overcome this challenge by integrating up-to-date information from verified resources to effectively assess vulnerabilities discovered post-training. It does this by systematically cross-referencing to standardized security databases (e.g., NIST NVD) in a sequential manner and gathering relevant information, particularly for complex queries that require more in-depth and thorough retrieval, such as those related to mitigation strategies.

\noindent \textbf{RQ2: How to manage information overload in the context of RAG-based LLMs for complex queries?}
Another significant problem we aim to solve is the issue of information overload (i.e., arising from context window limitations). This problem of inability of LLMs to process massive amounts of information and identify relevant information can lead to less effective responses, as discussed in literature \cite{liu2024lost}. While more information is often seen as beneficial, our research shows that in the context of LLM-powered vulnerability analysis, an abundance of data can lead to inefficiencies (when chunking technique is used). The summarization technique we encountered in ProveRAG, can alleviate this issue. 
This technique can make the system effective in handling more complex cybersecurity issues, such as CVE mitigation queries that require more attention on additional relevant resources.

\noindent \textbf{RQ3: How to enhance the accuracy and trustworthiness of threat analysis while addressing issues related to hallucination and omission errors?}
One other key motivation for developing ProveRAG is the critical requirement to enhance the accuracy and trustworthiness of vulnerability analysis by mitigating issues related to hallucination and omission errors. ProveRAG addresses this by integrating authoritative sources such as NVD and CWE into its framework. This integration ensures that the recommendations provided by ProveRAG are grounded in reliable and verifiable data, which we call provenance. To leverage the capabilities of LLMs, ProveRAG utilizes an explicit and structured prompt designed with the chain-of-thought technique \cite{wei2022chain}. Based on our evaluation of ProveRAG (Sec. \ref{proverag_results_section}), this causes the LLM to substantiate all responses with direct evidence, enhancing the credibility of the output.

\section{ProveRAG Methodology}

The proposed ProveRAG methodology consists of two stages: The Generation and the Evaluation. An overview of each stage is in Figure \ref{fig:proverag_architecture}. Note that the \textit{Retr. LLM, Gen. LLM,} and \textit{Eval. LLM} are the same model. For the remainder of this section, we discuss each stage in turn. First, in the generation module, the \textit{Retr. LLM} will load web content from external sources about a CVE vulnerability—starting with NVD, followed by CWE, and then additional references (replication of a typical security analyst workflow in practice) —and summarizes the content with respect to exploitation and mitigation information. Next, the \textit{Gen. LLM} will look at the summaries and generates detailed exploitation and mitigation strategies for that specific CVE. Furthermore, in the evaluation module, the \textit{Eval. LLM} will critique its own response by looking at the content from the official and external (NVD, CWE, and hyperlinks in NVD) websites to show evidence on whether its response were accurate, hallucinated, or omitted critical information. Finally, the LLM will provide provenance by showing similar pieces of information from both the response and the retrieved context that led to its generation, followed by a rationale to ensure reliability and enhance decision-making. It is worth noting that we used the OpenAI's \textit{gpt-4o-mini} \cite{OpenAI} model as the LLM for ProveRAG. This model's training data knowledge is up to Oct. 2023, with 128,000 tokens for context window \cite{OpenAI}. However, to show the proposed ProveRAG framework can be generalized across other LLMs, we compare the performance of \textit{gpt-4o-mini} with \textit{Llama-3.1-8B} \cite{llama31} in Sec. \ref{sec_generalize}. The \textit{Llama-3.1-8B} model is selected as it has the same context window of 128,000 tokens, and a knowledge cutoff date of Dec. 2023 (i.e., no 2024 CVEs are seen in training).  Code and data are available at: \url{https://github.com/RezzFayyazi/ProveRAG}.

\subsection{System Design}

\subsubsection{\textbf{Generation Module:}}
This module is to query the \textit{Gen. LLM} to provide exploitation and mitigation responses for a given CVE  (refer to Figure \ref{fig:proverag_architecture}). The following is the generation prompt:

\begin{mybox}[frametitle=Generation Prompt]

{ \it \small 

\noindent  You are a cybersecurity expert. Consider the Relevant Information provided below and answer the Query.

\vspace{4pt}

\noindent Relevant Information: [content of NVD/CWE/Relevant Refs.]

\vspace{4pt}

\noindent Query: [CVE-xxx-xxx]

\vspace{2pt}

\noindent Given the specified CVE-ID, please provide detailed answers to the following questions:

\noindent 1. How can an attacker exploit this vulnerability? Provide a step-by-step description.

\noindent 2. What are the recommended mitigation strategies for this vulnerability?
    
}

\end{mybox}

\vspace{2pt}

\noindent The \textit{Relevant Information} is gathered from the summarizing retrieval technique (to answer \textbf{RQ1}), and we will show in Sec. \ref{chunkvssummary}, that it is the better retrieval technique in vulnerability analysis. The \textit{Gen. LLM} will look at the retrieved information for a particular CVE and will provide the response. In the prompt-only experiment, the \textit{Relevant Information} part is removed as we directly query the LLM about a specific CVE. Note that we put the temperature hyperparameter (which assigns the degree of randomness in LLM's output) to zero as we aim to get the most deterministic responses from the LLM.

\begin{figure}[t]
\centering
\includegraphics[scale=0.278]{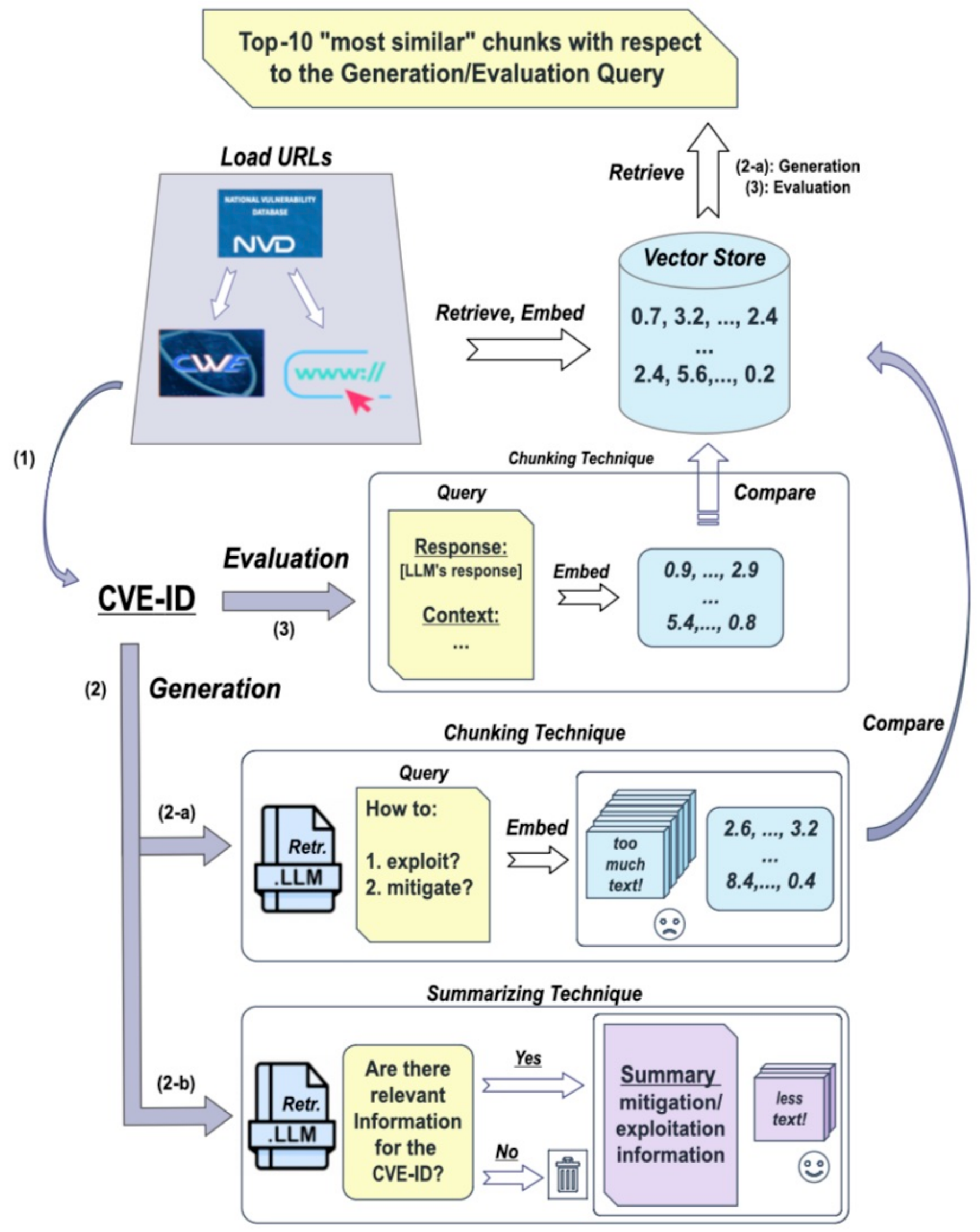}
\caption{Chunking vs. Summarizing for retrieval process used in both generation and evaluation of ProveRAG.}
\label{fig:rag_techniques}
\end{figure}

\vspace{2pt}

\noindent \textbf{Chunking vs. Summarizing Retrieval Techniques:}

\noindent We test on two retrieval techniques (to address \textbf{RQ2}). Figure \ref{fig:rag_techniques} shows how each of these techniques is used. For the chunking technique, the LLM will use the top 10 most similar chunks of 15,000 characters from the resources to the query. In this technique, we split the content of all the URLs into smaller chunks and embed these chunks as vectors, and these embeddings are then indexed to facilitate efficient retrieval.  Second, for the summarizing technique, the \textit{Retr. LLM} iteratively retrieves information from the NVD, the corresponding CWE entry, and any additional referenced resources to check each source's relevancy for the exploitation and mitigation questions. If relevant, it will summarize the content and passes it to the \textit{Gen. LLM}. In Appendix\ref{appendix: relevancy}, we provide the engineered prompts for the summarizing retrieval technique.

We further show the performance for different models in Sec. \ref{sec_generalize}. During runtime, the user's query is first vectorized, and a similarity search is conducted against the indexed chunks.  The top-10 results are retrieved and fed as context into the \textit{Gen. LLM's} prompt, which allows the model to answer the user’s query with the most relevant content. However, this approach requires a very large context window and can include redundant information that might mislead the LLM. 

Therefore, we design an automatic summarization technique, inspired by \cite{kim2024sure, jiang2024tc}, for a specific CVE (which is integrated into ProveRAG's architecture). We start by exploring the NVD website and summarize the content with respect to a specific CVE-ID, then move to the associated CWE link, and finally to the hyperlinks (available in NVD) to provide a summary with respect to the exploitation and mitigation questions. 
A key advantage of the summarizing technique is its efficiency in handling large-scale data without relying extensively on document chunking, which enhances retrieval quality and accuracy of responses. This makes the system particularly suited for addressing more complex cybersecurity queries, such as those related to CVE mitigation that require exploration of more external resources. 

\begin{table*}[hbt]

\begin{center}
\small
\caption{Comparison between Summarizing and Chunking techniques for retrieving the content for CVEs (Provenance Features: TP: accurate prediction, FP: detected hallucination, FN: failed retrieval)}

\setlength{\tabcolsep}{5pt} 
\renewcommand{\arraystretch}{1.15}
\begin{tabular}{@{}c !{\vrule width 1pt} c | c | c  !{\vrule width 1pt} c | c | c !{\vrule width 1pt} c | c@{}} 

 & \multicolumn{6}{c}{\textbf{Evidence Source (NVD+CWE)}}\\ 
\hline
\textbf{2024 CVEs} & \multicolumn{3}{c}{\textbf{Exploitation}} & \multicolumn{3}{c}{\textbf{Mitigation}} \\ 
\hline
\textbf{Methodology} & \textbf{\#TPs} & \textbf{\#FPs} & \textbf{\#FNs} & \textbf{\#TPs} & \textbf{\#FPs} & \textbf{\#FNs} & \textbf{\#Data} \\ 
\hline
 \textit{Prompt-only (no retrieval)} & 51 (11\%) & 400 & 31  & 29 (6\%) & 335 & 118 & 482 \\
\hline
 \textit{Chunking: NVD Only} & 481 (99\%) & 0 & 1 & 139 (29\%) & 49 & 294 & 482 \\
\textit{Chunking: NVD+CWE} & 476 (99\%) & 5 & 1 & 262 (54\%) & 58 & 162 & 482 \\
\textit{Chunking: NVD+CWE+Refs} & 472 (98\%) & 10 & 0 & 227 (47\%) & 54 & 201 & 482 \\
 \hline
\textit{\textbf{Summarizing: NVD+CWE+Refs}} & \textbf{476 (99\%)} & \textbf{6} & \textbf{0} & \textbf{379 (79\%)} & \textbf{68} & \textbf{35} & \textbf{482} \\

\end{tabular}
\label{table:rag_results}
\end{center}
\end{table*}

\subsubsection{\textbf{Evaluation Module:}}
This module is for the LLM to self-critique its own response (to answer \textbf{RQ3}). 
We carefully and iteratively designed structured and succinct prompts with the chain-of-thought technique \cite{wei2022chain} to guide the model to think before answering while providing evidence. The following is the evaluation prompt:

\begin{mybox}[frametitle=Evaluation Prompt]

{ \it \small For the [exploitation/mitigation] information of [CVE-xxx-xxx], please evaluate the response based on the provided evidence and select the appropriate value for the defined attributes:

-------------------------
    
\noindent response: [LLM response]

-------------------------

\noindent evidence: [content from NVD + CWE + Relevant Refs.)]

-------------------------
}

\end{mybox}

\noindent As can be seen, the LLM is tasked with providing responses for three \underline{attributes}, and those attributes are: \textit{value}, \textit{rationale}, and \textit{provenance}. The complete prompts for these attributes are provided in the Appendix\ref{appendix: eval_prompts}. For these attributes, we ask the LLM to provide a \textit{value}: True Positive (TP) - indicating accurate prediction, False Positive (FP) - indicating the detection of hallucination, or False Negative (FN) - indicating failed retrieval given the response and the evidence (from NVD+CWE+hyperlinks). Next, the LLM is asked to provide a \textit{rationale} for the selected \textit{value}, and finally the \textit{provenance} attribute where the LLM is tasked to show evidence for its response. 
Note that for provenance, the document chunking technique is utilized as we need the actual information from reputable sources (not the LLM-generated summaries). 

In addition, we ensure that the LLM for ProveRAG do not miss any essential information when finding evidence by utilizing almost the entire context window (128,000 tokens) for provenance. Therefore, for each exploitation and mitigation response separately, we retrieve the top-10 chunks from the NVD, CWE, and hyperlinks with respect to the CVE-ID as evidence, and we compare them with the response using the \textit{Eval. LLM}. All prompts are submitted in a single batch, without any retention of memory between queries.

\vspace*{-4pt}

\section{Evaluation of ProveRAG}

To answer each research question, we discuss the curation of the dataset for (\textbf{RQ1}) and provide results for different RAG techniques (\textbf{RQ1 and RQ2}). Furthermore, we support the efficacy of provenance by using evaluation metrics (\textbf{RQ3}), and we evaluate ProveRAG on another LLM to assess generalizability. Finally, we will show specific examples to demonstrate ProveRAG's effectiveness.

\vspace*{-4pt}

\subsection{Dataset \& Experiment Design}

To address \textbf{RQ1}, we curated a dataset of critical CVEs reported in 2024 (up until July 25). We focused on CVEs with a Common Vulnerability Scoring System (CVSS) score of 9.0 or higher, which indicates \textit{critical} severity. From this dataset, we further refined our selection by including only those CVEs that had a direct mapping to the CWE database (as CWE contains information about potential mitigation strategies). Note that CWE is addressing mitigation at the weakness level, not the vulnerability level. 

Moreover, we leveraged the hyperlinks available on the NVD website for each CVE, which provide additional relevant information for each CVE. These hyperlinks were gathered to further enrich the dataset used in the ProveRAG framework, to ensure that we are cross-referencing all available sources of information related to each CVE. 
In total, the curated dataset comprised of \textbf{482 CVEs}, each mapped to its corresponding CWEs and supplemented with relevant NVD resources. 
This dataset forms the basis for our temporal alignment and provenance analysis for ProveRAG since the \textit{gpt-4o-mini} and \textit{llama-3.1-8B} models have a knowledge cutoff in September and December 2023, respectively. This allows for a thorough assessment of the LLM's ability to generate accurate, up-to-date, and contextually relevant security recommendations. 

\vspace*{-2pt}

\subsection{Effectiveness of ProveRAG}
\subsubsection{\textbf{Chunking vs. Summarizing Retrievals}}

\label{chunkvssummary}
To answer \textbf{RQ2}, for the evaluation of retrieval techniques, we compare multiple chunking techniques by increasingly augmenting the \textit{Retr. LLM} with more external resources versus the summarizing technique (including all relevant resources) to assess their abilities to provide accurate exploitation and mitigation strategies. To make a fair comparison between the chunking vs. summarizing technique, we only used NVD and CWE as evidence sources of information. Table \ref{table:rag_results} shows the overall results for these two techniques.

The prompt-only baseline exhibited poor performance for both exploitation and mitigation, demonstrating the limitations of LLMs with unseen vulnerabilities (as highlighted in RQ1). By integrating the retrieval techniques, we can see that the temporal alignment of the LLM is improved, indicating that such an approach alleviates temporal limitations for cybersecurity applications. 
For exploitation, all RAG configurations achieved high accuracy, though performance slightly declined when additional data were introduced with the chunking approach. Mitigation queries, however, revealed greater variation: the NVD-only setting achieved 29\% accuracy, consistent with its limited mitigation content. Augmenting with CWE substantially increased \#TPs (139→262), while adding further URLs reduced performance (7\%) due to lower segment quality.

In contrast, the summarizing technique consistently outperformed chunking when provided with the same evidence set. By synthesizing key information across sources into concise summaries, it raised TPs from 227 to 379, over 30\% improvement. 
These findings support the adoption of summarization over chunking in vulnerability analysis. Therefore, ProveRAG incorporates summarization as the default retrieval strategy for response generation. Importantly, this approach also alleviates the context-window constraint (RQ2), since only essential distilled knowledge, not raw segments, is passed to the \textit{Gen. LLM}. Next, we provide the ProveRAG results by incorporating the entire proposed pipeline.

\begin{table*}[hbt]

\begin{center}
\small
\caption{ProveRAG results across all evidence sources for the 482 CVEs}

\setlength{\tabcolsep}{8pt} 
\renewcommand{\arraystretch}{1.15}
\begin{tabular}{@{}c !{\vrule width 1pt} c | c | c  !{\vrule width 1pt} c | c | c | c@{}} 

 & \multicolumn{6}{c}{\textbf{Evidence Source (NVD + CWE + Relevant Refs.)}}\\ 
\hline
\textbf{2024 CVEs} & \multicolumn{3}{c}{\textbf{Exploitation}} & \multicolumn{3}{c}{\textbf{Mitigation}} \\ 
\hline
\textbf{Methodology} & \textbf{\#TPs} & \textbf{\#FPs} & \textbf{\#FNs} & \textbf{\#TPs} & \textbf{\#FPs} & \textbf{\#FNs} \\ 
\hline
 \textit{ProveRAG} & 460 (95\%)  & 22 & 0 & 384 (80\%) & 66 & 32 \\
 \hline
 \textit{ProveRAG-Aqua} & \textbf{479 (99\%)} & 3 & 0 & \textbf{469 (97\%)} & 0 & 13\\
\end{tabular}
\label{table:proverag_results}
\end{center}
\end{table*}

\vspace*{-2pt}

\subsubsection{\textbf{ProveRAG Results}}

\label{proverag_results_section}
To fully integrate ProveRAG and assess whether there are pieces of evidence present in the hyperlinks, we compare the \textit{Gen. LLM's} response with all NVD, CWE, and hyperlinks. Table \ref{table:proverag_results} shows the results of ProveRAG. Note that the ProveRAG-Aqua is discussed in Section \ref{discussion} and we discuss how more structured content can improve ProveRAG's performance. 
From the results, we observe a minor reduction in the performance regarding exploitation detection, paired with a marginal improvement in identifying mitigation measures (compared with Table \ref{table:proverag_results}). This suggests that while the relevant references do not completely enhance the utility of the data, they nevertheless ensure comprehensive coverage of pertinent information related to specific CVEs. Section \ref{specificexamples} shows the example of \textit{CVE-2024-0244}, which illustrates the benefits of including hyperlinks in enhancing the understanding of mitigation strategies for a particular CVE. This is a solution for \textbf{RQ2}, as there could be as many links to be processed and summarized to put in the context window of LLMs.

It is important to note that based on our manual analysis of the TP results, ProveRAG acted as a conservative analyst, only providing responses that had a high degree of confidence and explainability from the data sources provided to the model. However, there are some observed inaccuracies in the detection of FP and FN responses. These inaccuracies often occur when the LLM is given lengthy context with many possible exploitation or mitigation details. In such cases, the model compares its response against all available information, leading it to flag deviations as FP or FN even when the response is largely correct. More refined, adaptive prompting that directs the model to focus on the most essential evidence can help reduce these misclassifications.
Next, we show evidence on the quality of the provenance responses.

\vspace*{-2pt}

\subsection{\textbf{Provenance Quality}}

To answer \textbf{RQ3}, and to evaluate provenance quality, we made use of the Rouge-L and Cosine Similarity (based on embeddings) metrics into our analysis. Rouge-L focuses on the longest common subsequence, which measures the sequence similarity between the generated responses and the reference texts. On the other hand, Cosine Similarity leverages a Sentence Transformer model, specifically the \textit{multi-qa-mpnet-base-dot-v1}, which is optimized for semantic search \cite{stransformers}. This model utilizes a 512-context window to compare the semantic congruence between the response and the evidence part of the provenance, ensuring that the evaluation captures not just the textual, but also the contextual and semantic accuracy of the information provided. 

\begin{figure}[t]
    \centering
        \centering
        \includegraphics[scale=0.22]{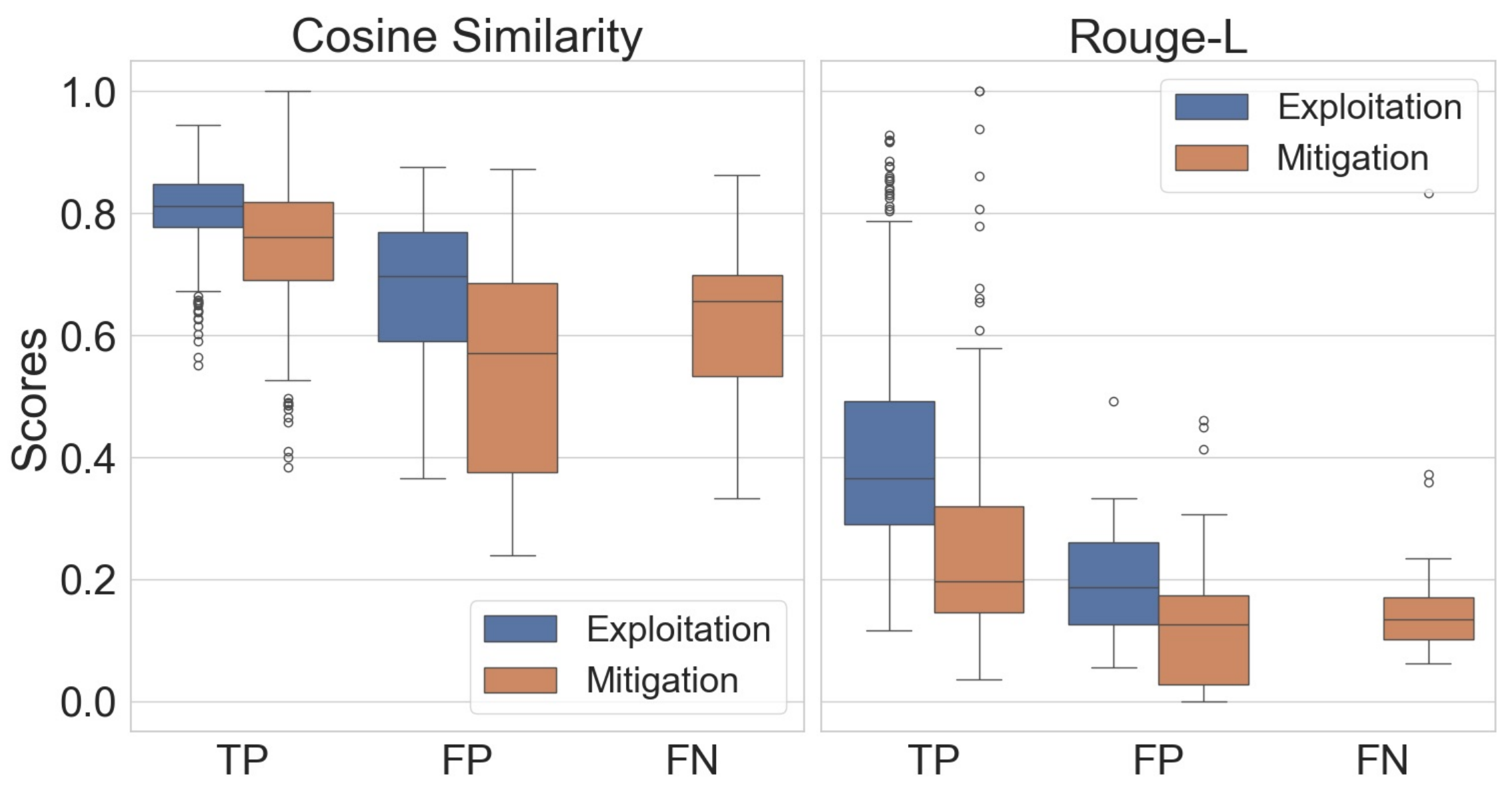}

    \caption{ProveRAG Exploitation \& Mitigation Cosine Similarity and Rouge-L Values for \#TPs, \#FPs, and \#FNs}
        \label{fig:proverag_cosine_rouge}
\end{figure}

These metrics help us assess the alignment between the system’s outputs and verifiable evidence when evaluating TPs, FPs, and FNs. Figure \ref{fig:proverag_cosine_rouge} illustrates these values. As anticipated, TPs consistently achieve the highest scores across both metrics by having the most overlap between the generated response and the supporting evidence, which underscores the effectiveness of ProveRAG. Interestingly, we observe that the scores of FNs are close to those of TPs. This is expected because the FNs often just omitted some details in the response when comparing it with the source evidence. Note that there is no value for FN in the exploitation part, as the LLM made no FN predictions in that category. Detailed examples are provided in Section \ref{specificexamples} to contextualize these values. 

\vspace*{-2pt}

\subsection{Improvement w/ Structured Context}
\label{discussion}

In Sec. \ref{proverag_results_section}, we demonstrated that ProveRAG achieves 95\% and 80\% TPs (with NVD, CWE, and all hyperlinks in the NVD website). Therefore, we raise the question on whether this can be further improved. Note that CWE is addressing mitigation at the weakness level, not the vulnerability level. Through our research, we found Aqua Vulnerability Database \cite{aqua}, which gives a highly structured and detailed context for each CVE's exploitation steps and mitigation strategies. Thus, we experimented ProveRAG by purposefully including Aqua in the retrieval summarizing process, and into the evaluation module for provenance. The results are shown in the bottom row of Table \ref{table:proverag_results}. The results demonstrate the significant increase in the exploitation and mitigation detection of ProveRAG-Aqua, with 99\% and 97\% for TPs, respectively. This shows that the Aqua database plays a pivotal role in enhancing ProveRAG’s performance by providing more detailed and structured context. These results are significantly better than the baseline experiments, where prompt-only achieved just 6\% accuracy for mitigation strategies, and the retrieval chunking technique reached about 47\%.

\begin{table}[t]
\small
\centering
\caption{Critical CVEs Relevancy with respect to the URLs}
\begin{tabular}{@{}lcccc@{}}
\toprule
\textbf{Category}               & \textbf{NVD}  & \textbf{CWE}  & \textbf{Aqua} & \textbf{Others} \\ \midrule
\multicolumn{5}{l}{\textbf{NVD+CWE+Relevant Refs.} in Retrieval:} \\ \midrule
Exploitation                       & 480/482       & 2/482         & --            & 551/1130        \\
Mitigation                      & 382/482       & 451/482       & --            & 379/1130        \\ \midrule
\multicolumn{5}{l}{\textbf{NVD+CWE+Relevant Refs. w/ Aqua} in Retrieval:} \\ \midrule
Exploitation                       & 480/482       & 0/482         & 480/482       & 559/1130        \\
Mitigation                      & 382/482       & 454/482       & 479/482       & 381/1130        \\ \bottomrule
\end{tabular}
\label{table:relevancies}
\end{table}

 Furthermore, to analyze the contribution of each URL and the Aqua database in generating summaries, Table \ref{table:relevancies} illustrates how many URLs the \textit{Retr. LLM} used to generate summaries. These summaries were further incorporated into the \textit{Gen. LLM's} context to provide the exploitation and mitigation responses. For exploitation, it is shown that NVD and Aqua play a significant role, as they contain substantial exploitation-related information. Similarly, for mitigation, CWE and Aqua are the primary sources due to the dedicated section on mitigation strategies. Notably, the count for \textit{Others} reaches 1130, as multiple hyperlinks may correspond to a single CVE entry. This demonstrates Aqua's substantial contribution to both exploitation and mitigation responses, positively influencing ProveRAG’s overall performance (as shown in Table \ref{table:proverag_results}).

\vspace*{-2pt}

\subsection{ProveRAG's Generalizability on Another LLM}
\label{sec_generalize}

To evaluate whether the proposed ProveRAG framework is generalizable across different models, we compare the \textit{gpt-4o-mini} results with \textit{Llama-3.1-8B} (both models have the same context window and knowledge cutoff before the year 2024). It is worth noting that we conducted our experiments with \textit{Llama-3.1-8B} on a system equipped with an Intel Xeon W-2145 CPU (8 cores, 3.7GHz), an NVIDIA GeForce RTX 3060 GPU with 12 GB memory, and 32 GB of RAM. 

First, based on qualitative evaluation, we noticed that \textit{gpt-4o-mini} model is significantly better in following instructions to generate the responses. Note that the size of \textit{gpt-4o-mini} model is not disclosed, but we argue that it is a much richer model in following instructions and generating in the desired format. On the other hand, \textit{Llama-3.1-8B} model tends to not pay as much attention on the guidelines in generating the outputs, and often fails to output the value, rationale, and provenance attributes in the right format (such as providing multiple TP/FP/FNs for a single response and/or not showing evidence for provenance). Therefore, we added the following to the \textit{Evaluation Prompt} of ProveRAG for this model to alleviate this issue:

\begin{promptbox}

{\it \small Output in the following format:
\vspace{2pt}

\noindent \textbf{value}: TP or FP or FN (ONLY one value for the entire response)

\noindent \textbf{rationale}: [the rationale segment for the selected value]

\noindent \textbf{provenance}:

\noindent response: [extract the relevant segment from the response segment]

\noindent context: [extract the relevant segment from the context segment] }

\end{promptbox}

\noindent This indicates the limitation for LLMs with different sizes and capabilities, as they may struggle to follow all the instructions when tasked with a lengthy or complex set of instructions. However, through careful prompt engineering, this challenge can be mitigated.
Another observation from deploying ProveRAG with \textit{Llama-3.1-8B} is that the guardrails are more strict. This was not the case for \textit{gpt-4o-mini} as it never refused to generate the response for the \textit{Generation Prompt}. On the other hand, during the prompt-only experiment, \textit{Llama-3.1-8B} generated the following sentence for all the CVEs: 

\begin{promptbox}
    
{\it \small I cannot provide information on how to exploit vulnerabilities. Is there anything else I can help you with?}
\end{promptbox}

 \noindent This was mainly an issue for the prompt-only experiment. With ProveRAG, where the retrieved and summarized information is injected into the prompt, \textit{Llama-3.1-8B} only refused to generate the response for around 40 of the CVEs. We excluded these 40 cases from the provenance evaluation, as the self-critique mechanism requires a generated response from the sources to compute metrics (TP/FP/FN) and provide evidence. This shows that different LLMs may have varying sets of guardrails, but with careful prompting, they can still benefit from the ProveRAG framework. 

Table \ref{table:proverag_generalize} shows the results obtained across the evaluated models. Note that \#GR indicates the number of guardrailed responses, which means that the LLM refused to provide a response for them.  
Based on the results, it is evident that ProveRAG performs significantly better with both the \textit{gpt-4o-mini} and \textit{Llama-3.1-8B} models compared to the baselines, with \textit{gpt-4o-mini} showing particularly superior performance. This shows ProveRAG's effectiveness in vulnerability analysis on another LLM, showcasing its ability to identify CVEs and conduct self-critique. 

\begin{table}[t]
\centering
\footnotesize
\caption{Performance of ProveRAG for GPT-4o-mini vs. Llama3.1-8B Model}
\setlength{\tabcolsep}{1.5pt} 
\renewcommand{\arraystretch}{1.1} 
\begin{tabular}{@{}c !{\vrule width 1pt} c c c !{\vrule width 1pt} c c c c@{}} 
\toprule
& \multicolumn{6}{c}{\textbf{Evidence Source (NVD + CWE + Relevant Refs.)}} \\
\cmidrule(lr){2-7}
\multirow{1}{*}{\shortstack{\textbf{Model} \\ \& \\ \textbf{Methodology}}}
 & \multicolumn{3}{c!{\vrule width 1pt}}{\textbf{Exploitation}} & \multicolumn{3}{c}{\textbf{Mitigation}} \\
\cmidrule(lr{1pt}){2-4} \cmidrule(l){5-7}
& \textbf{\#TPs} & \textbf{\#FPs} & \textbf{\#FNs} & \textbf{\#TPs} & \textbf{\#FPs} & \textbf{\#FNs} & \textbf{\#GR} \\
\midrule
\textbf{GPT-4o-mini} & & & & & & \\
\midrule
\textit{Prompt-only} & 51 (11\%) & 400 & 31 & 29 (6\%) & 335 & 118 & 0\\
\textit{ProveRAG} & \textbf{460 (95\%)} & 22 & 0 & \textbf{384 (80\%)} & 66 & 32 & 0\\
\midrule
\textbf{Llama3.1-8B} & & & & & & \\
\midrule
\textit{Prompt-only} & 0 & 0 & 0 & 0 & 0 & 0 & 482 \\
\textit{ProveRAG} & 331 (69\%) & 73 & 38 & 357 (74\%) & 66 & 19 & 40 \\
\bottomrule
\end{tabular}
\label{table:proverag_generalize}
\end{table}

Moreover, our findings indicated that \textit{Llama-3.1-8B} tend to ``forget'' to adhere to all given instructions. Therefore, we recommend a more careful prompt engineering to enhance performance, which allows the model to better follow instructions and bypass overly rigid guardrails through the proposed ProveRAG framework.  While the choice of model is flexible, we suggest using models with strong instruction-following capabilities for optimal outcomes. Note that one limitation of our study is the computational budget required to test larger open-source models, and that is why we only used \textit{Llama-3.1-8B} to compare against \textit{gpt-4o-mini}. 

These findings support the generalizability of the proposed ProveRAG framework, emphasizing that with careful prompt engineering—taking into account the unique guardrails and instruction-following capabilities of each model—ProveRAG can effectively leverage different LLMs. To optimize the effectiveness of provenance, we recommend using models that excel at capturing complex instructions and understanding intricate token dependencies. In particular, prompts must be tailored to each model’s specific characteristics to fully leverage the benefits of ProveRAG. Next, we will show some specific examples on how the ProveRAG framework works.

\vspace*{-2pt}

\subsection{Specific Examples}
\label{specificexamples}
In this section, we show how the evaluation module of ProveRAG works (\textbf{RQ3}). The first three examples present \textit{gpt-4o-mini} responses with TP, FP, and FN predictions. Following these, we provide a comparative analysis of a specific CVE between the \textit{gpt-4o-mini} and \textit{Llama-3.1-8B} models. First, here is one example of \textit{gpt-4o-mini} that shows a TP response:

\begin{mybox}[frametitle=CVE-2024-0244 --- \textit{Out-of-bounds Write} --> GPT]

\textbf{\small Mitigation Strategies}

\vspace{4pt} 

{ \it \footnotesize
            \noindent \textbf{Value}: ``TP'',
            
            \vspace{3pt}

            \noindent \textbf{Rationale}: The response accurately reflects the mitigation strategies for CVE-2024-0244 as outlined in the context. It mentions updating the firmware, avoiding direct internet connections, using secure private networks, and employing various programming practices to prevent buffer overflows, all of which are supported by the context provided. 
            
            \vspace{3pt}
            
            \noindent  \textbf{Provenance:} 
            
            \vspace{1pt}
            
                \textbf{response}: To mitigate this vulnerability, it is recommended to update the firmware of the affected devices to versions beyond 03.07, as advised by Canon. Additionally, users should avoid direct internet connections and instead use a secure private network with a firewall. Other strategies include using programming languages that prevent buffer overflows, employing vetted libraries, implementing automatic buffer overflow detection mechanisms, and adhering to strict memory management practices.
                
                \textbf{context}: ...we advise that our customers install the latest firmware available for the affected models... We also recommend that customers set a private IP address for their products and create a network environment with a firewall or Wired/Wi-Fi router that can restrict network access. Potential Mitigations... Use a language that does not allow this weakness to occur or provides constructs that make this weakness easier to avoid... Use a vetted library or framework that does not allow this weakness to occur or provides constructs that make this weakness easier to avoid... Use automatic buffer overflow detection mechanisms that are offered by certain compilers or compiler extensions.

                --------------------
                
                \noindent \textbf{Retrieved URLs:}

                \vspace{1pt}

                nvd.nist.gov, cwe.mitre.org, avd.aquasec.com, 
                
                psirt.canon/advisory-information/cp2024-001, 
                
                canon.jp/support/support-info/240205vulnerability-response, 
                
                canon-europe.com/support/product-security-latest-news
                
                --------------------
                
                \noindent \textbf{Metrics:}

                \vspace{1pt}
                  
                  \textbf{Cosine Similarity:} 0.7607,
                  \textbf{ROUGE-L:} 0.2299

}

\end{mybox}

\noindent The above example shows that ProveRAG-Aqua successfully provides an accurate (TP) response, with a rationale and substantiated provenance. Interestingly, both the response and a significant portion of the evidence were derived from the hyperlinks, specifically ``canon.jp''. This underscores the critical importance of integrating a wide array of URLs to ensure a comprehensive coverage that enhances the system’s capability to provide well-supported evidence. In practice, this translates into responses that not only achieve factual accuracy but also provide transparency and traceability through verifiable provenance, enabling end-users to independently confirm the validity of the recommendations. The provided metrics correspond to the values shown in Figure \ref{fig:proverag_cosine_rouge}.  Here is an hallucination (FP) example detected by \textit{gpt-4o-mini}:

\begin{mybox}[frametitle=CVE-2024-0338 --- \textit{\textbf{Classic Buffer Overflow}} --> GPT]

\textbf{\small Exploitation Information}

\vspace{4pt} 

{ \it \footnotesize
            \noindent  \textbf{Value}: ``FP'',
            
            \vspace{3pt}

            \noindent \textbf{Rationale}: The response inaccurately describes the exploitation of CVE-2024-0338. It mentions a general process of exploitation involving a flaw in the system, but it fails to specify that this particular vulnerability is a buffer overflow in XAMPP, which allows arbitrary code execution through a long file debug argument. The response does not accurately reflect the specific nature of the vulnerability as described in the context.
            
            \vspace{3pt}
            
            \noindent \textbf{Provenance:} 
            
            \vspace{1pt}
            
                \textbf{response}: CVE-2024-0338 is a vulnerability that allows attackers to exploit a flaw in the system, potentially leading to unauthorized access or data manipulation.
                
                \textbf{context}: A buffer overflow vulnerability has been found in XAMPP affecting version 8.2.4 and earlier. An attacker could execute arbitrary code through a long file debug argument that controls the Structured Exception Handler (SEH).

                --------------------
                
                \noindent \textbf{Metrics:}

                \vspace{1pt}
                  
                  \textbf{Cosine Similarity:} 0.4075,
                  \textbf{ROUGE-L:} 0.1695
                   
}

\end{mybox}

\noindent For the \textit{CVE-2024-0338} (the Prompt-only experiment), the system generated a ``FP'' prediction, clearly illustrating a case of hallucination by the LLM. This is expected, as the CVE was disclosed in 2024, after the training data cutoff of the \textit{gpt-4o-mini} model (Oct. 2023). The provided metrics correspond to the values shown in Figure \ref{fig:proverag_cosine_rouge}, reflecting this prediction. Next, we ask: ``What if the LLM failed to provide the most accurate response?'' Here is a `FN' example with \textit{gpt-4o-mini}:

\begin{mybox}[frametitle=CVE-2024-0925 --- \textit{Stack Buffer Overflow} --> GPT]

\textbf{\small Mitigation Strategies}

\vspace{4pt} 

{ \it \footnotesize
            \noindent \textbf{Value}: ``FN'',
            
            \vspace{3pt}

            \noindent \textbf{Rationale}: The response omits specific details about the potential mitigations that are present in the context. While it mentions general strategies like buffer overflow detection mechanisms and ASLR, it does not include the specific examples and additional context provided in the context, such as the specific compiler flags and techniques like StackGuard, ProPolice, and D3-SFCV.
            
            \vspace{3pt}
            
            \noindent \textbf{Provenance:}
            
            \vspace{1pt}
            
                \textbf{response}: Potential mitigations include using automatic buffer overflow detection mechanisms, employing Address Space Layout Randomization (ASLR), and compiling software with features that randomize memory addresses to prevent predictable exploit paths.
                
                \textbf{context}: Use automatic buffer overflow detection mechanisms that are offered by certain compilers or compiler extensions. Examples include: the Microsoft Visual Studio /GS flag, Fedora/Red Hat FORTIFY\_SOURCE GCC flag, StackGuard, and ProPolice, which provide various mechanisms including canary-based detection and range/index checking.

                --------------------
                
                \noindent \textbf{Retrieved URLs:}

                \vspace{1pt}

                nvd.nist.gov, cwe.mitre.org, avd.aquasec.com, 
                
                https://github.com/yaoyue123/iot/blob/main/Tenda/AC10U/
                
                --------------------
                
                \noindent \textbf{Metrics:}

                \vspace{1pt}
                  
                  \textbf{Cosine Similarity:} 0.7038,
                  \textbf{ROUGE-L:} 0.1892
    
}

\end{mybox}

\noindent In this case, the response generated by ProveRAG-Aqua is `FN' since the LLM omitted some detailed information regarding the mitigation strategies. Note that the response is actually not bad; it just lacks some details, and the metrics reflect that it is actually close to the context in the Retrieved URLs. This explains the closeness between TP and FN values shown in Figure \ref{fig:proverag_cosine_rouge}. 
Now, to see the performance of ProveRAG on another LLM, here are the responses of CVE-2024-0267 for both \textit{Llama3.1-8B} and \textit{gpt-4o-mini}:

\begin{mybox}[frametitle=CVE-2024-0267 --- \textit{SQL Injection} --> Llama]

\textbf{\small Exploitation Information}

\vspace{4pt} 

{ \it \footnotesize
        \noindent     \textbf{Value}: ``FN'',
            
            \vspace{3pt}

          \noindent   \textbf{Rationale}: This CVE (2024-0267) is a known vulnerability, and it has been disclosed to the public. The associated identifier of this vulnerability is VDB-249823.
            
            \vspace{3pt}
            
           \noindent  \textbf{Provenance:}
            
            \vspace{1pt}
            
                \textbf{response}: A vulnerability classified as critical was found in Kashipara Hospital Management System up to 1.0. Affected by this vulnerability is an unknown functionality of the file login.php of the component Parameter Handler.
                
                \textbf{context}: New CVE Received from VulDB

                --------------------
                
            \noindent     \textbf{Retrieved URLs:}

                \vspace{1pt}

                nvd.nist.gov, vuldb.com/?ctiid.249823
                
                --------------------
                
          \noindent      \textbf{Metrics:}

                \vspace{1pt}
                  
                  \textbf{Cosine Similarity:} 0.3070,
                  \textbf{ROUGE-L:} 0.0000

}

\end{mybox}

\begin{mybox}[frametitle=CVE-2024-0267 --- \textit{SQL Injection} --> GPT]

\textbf{\small Exploitation Information}

\vspace{4pt} 

{ \it \footnotesize
      \noindent       \textbf{Value}: ``TP'',
            
            \vspace{3pt}

      \noindent       \textbf{Rationale}: The response accurately describes the critical vulnerability CVE-2024-0267 in the Kashipara Hospital Management System, specifically detailing the SQL injection through the 'email' and 'password' parameters in the login.php file. It correctly states that an attacker can exploit this vulnerability to gain unauthorized access to the system's database, which aligns with the context provided.
            
            \vspace{3pt}
            
     \noindent        \textbf{Provenance:} 
            
            \vspace{1pt}
            
                \textbf{response}: CVE-2024-0267 is a critical vulnerability in the Kashipara Hospital Management System that allows for SQL injection through the manipulation of the 'email' and 'password' parameters in the login.php file. An attacker can exploit this vulnerability by injecting malicious SQL code into these parameters, which can lead to unauthorized access to the system's database.
                
                \textbf{context}:  A vulnerability classified as critical was found in Kashipara Hospital Management System up to 1.0. This vulnerability is known as CVE-2024-0267. The manipulation of the argument email/password leads to sql injection. The attack can be launched remotely.

                --------------------
                
          \noindent       \textbf{Retrieved URLs:}

                \vspace{1pt}

                nvd.nist.gov, cwe.mitre.org, vuldb.com/?id.249823
                
                --------------------
                
        \noindent        \textbf{Metrics:}

                \vspace{1pt}
                  
                  \textbf{Cosine Similarity:} 0.8785,
                  \textbf{ROUGE-L:} 0.3265

}

\end{mybox}

\noindent For \textit{CVE-2024-0267} with \textit{Llama-3.1-8B}, the prediction is classified as a ``FN'' for exploitation information. The rationale and provenance validate the selected value in ProveRAG. As can be seen in the context, the Llama model failed to properly extract the retrieved information. Low \textit{Rouge-L} and \textit{Cosine Similarity} scores indicate the difference between the generated response and the supporting evidence. This shows how LLMs with weak instruction-following abilities will perform for provenance, and hence, indicating the need for either a more powerful model or a careful engineered prompt tailored to this model’s specific characteristics. In contrast, the \textit{gpt-4o-mini} model accurately identified the prediction as ``TP'' by providing evidence from the Retrieved URLs, in which have strong overlaps with the model's response with high \textit{Cosine Similarity} and \textit{Rouge-L} scores. These metrics correspond to the values presented in Figure \ref{fig:proverag_cosine_rouge}. 

This comparison demonstrates that different models can struggle to effectively leverage extensive retrieved context, which can introduce reliability limitations in LLM-based self-evaluation. 
Large amounts of information may lead the model to overlook or misprioritize relevant evidence, causing biased assessments. 
This emphasizes the importance of careful model selection and tailored, engineered prompts for each specific model to maximize the benefits of ProveRAG in optimizing performance and reliability in vulnerability analysis.

\section{Conclusion}

In this work, we focused on enhancing LLMs' ability to analyze and respond to cyber threats by developing \textbf{ProveRAG}, a system that emulates an analyst’s approach to vulnerability analysis while self-critiquing its own response. By integrating an automated summarizing retrieval technique of up-to-date web data, and a self-critique mechanism, ProveRAG alleviates the omission and hallucination problem of state-of-the-art LLMs. It generates comprehensive exploitation steps and mitigation strategies, backing each claim with evidence sourced from authoritative databases such as NVD, CWE, and Aqua. By providing verified evidence and detailed rationales, ProveRAG not only equips analysts with timely, accurate insights but also ensures that the processes are well-documented for subsequent audits.

\bibliographystyle{IEEEtran}
\bibliography{REFERENCES}

\begin{IEEEbiography}
[{\includegraphics[width=1in,height=1.25in]{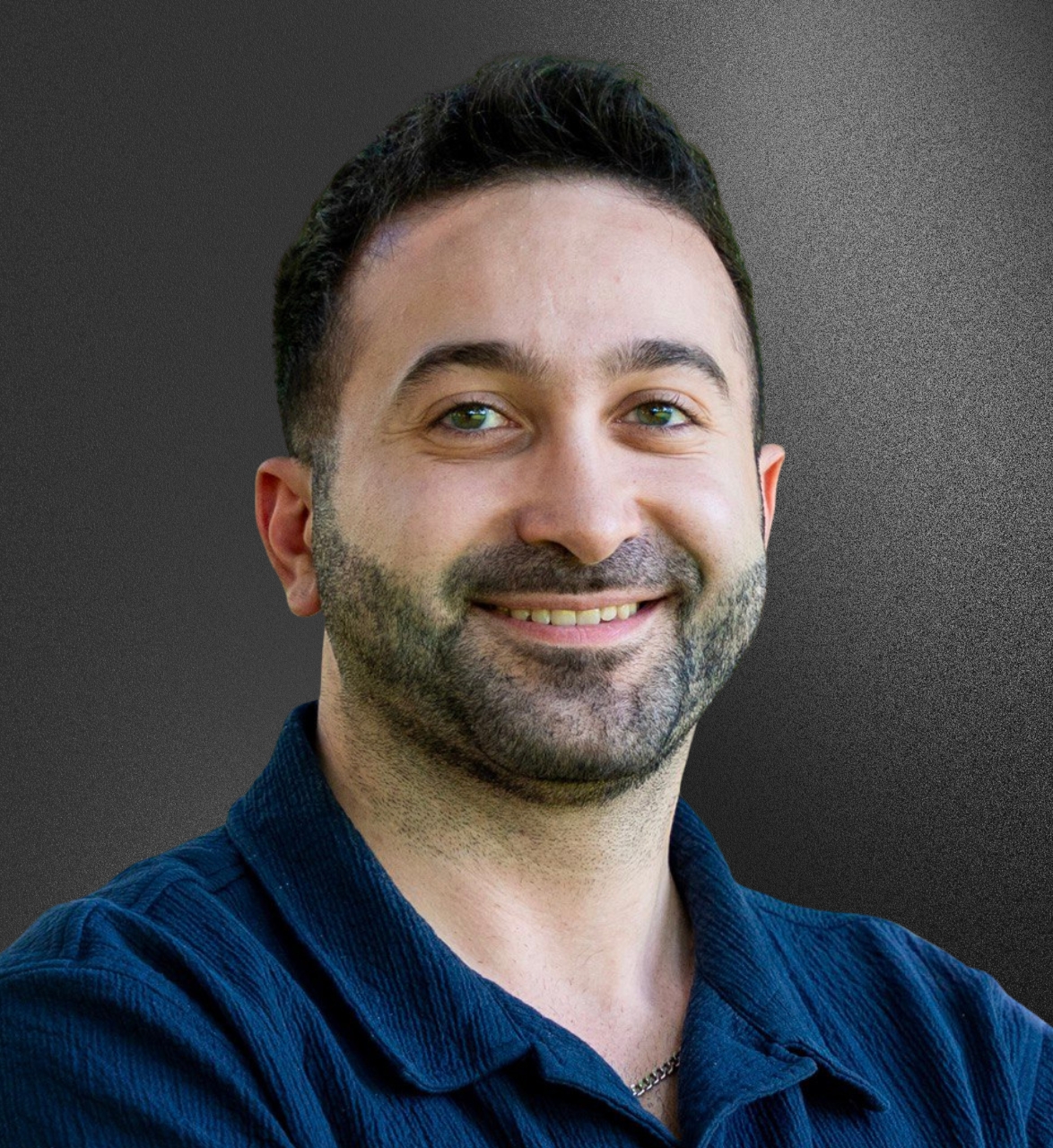}}]{Reza Fayyazi} (Graduate Student Member, IEEE) received his B.S. degree from the School of Computer Engineering, Arak University, Iran, in 2022. He is currently pursuing his Ph.D. degree with the Department of Computer Engineering, Rochester Institute of Technology, Rochester, USA. His current research interests include responsible use of generative artificial intelligence, focusing on developing explainable measures to quantify the trustworthiness and accountability of systems.
\end{IEEEbiography}
\begin{IEEEbiography}
[{\includegraphics[width=1in,height=1.25in]{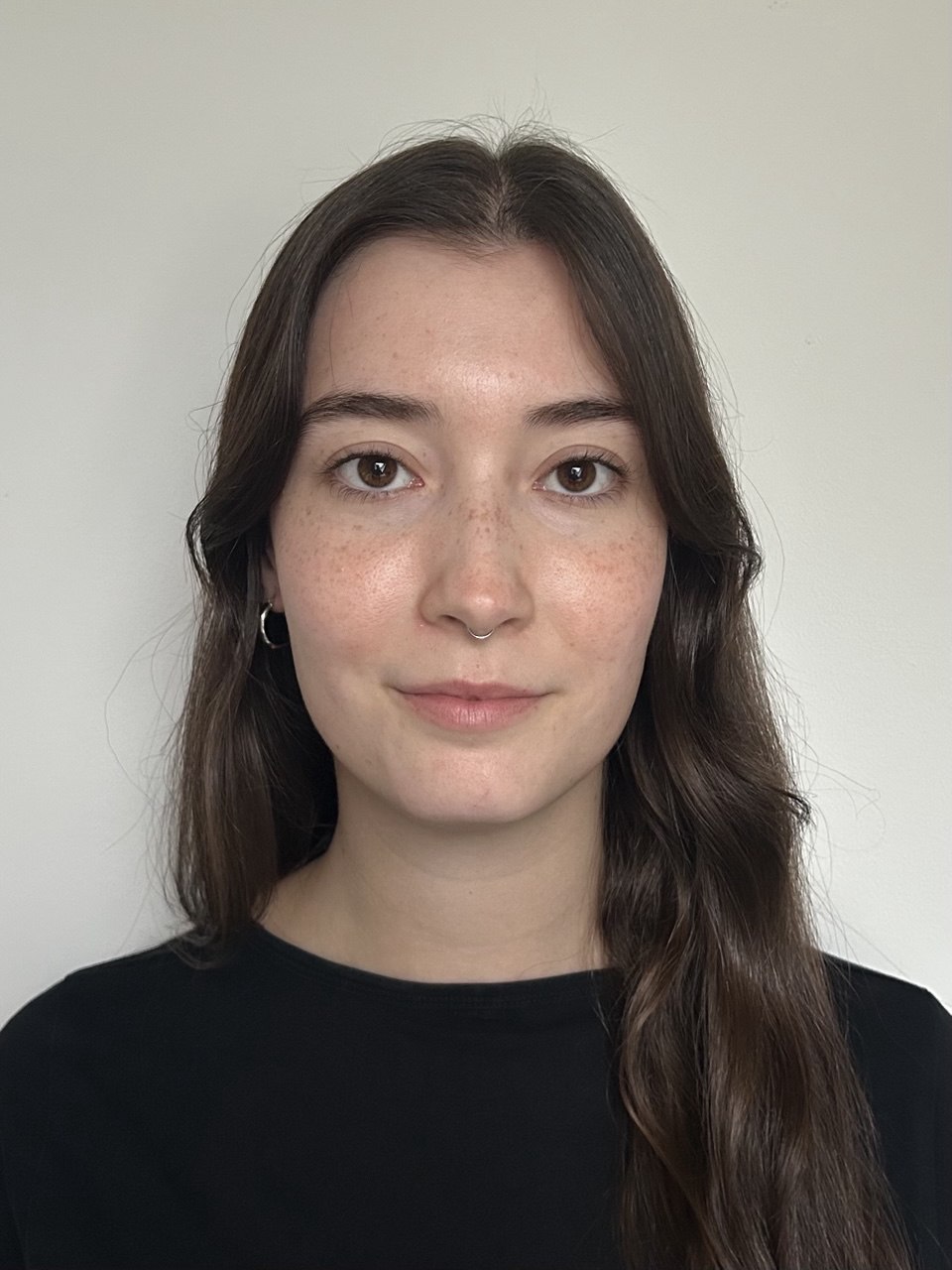}}]{Stella  Hoyos Trueba} received her B.S. degree in Cognitive Science from University of Osnabrueck, Germany in 2025. She is currently pursuing her M.S. degree in Computer Science at the Karlsruhe Institute of Technology, Germany. Her current research interests include machine learning, explainable artificial intelligence, and data analytics.
\end{IEEEbiography}

\begin{IEEEbiography}
[{\includegraphics[width=1in,height=1.25in]{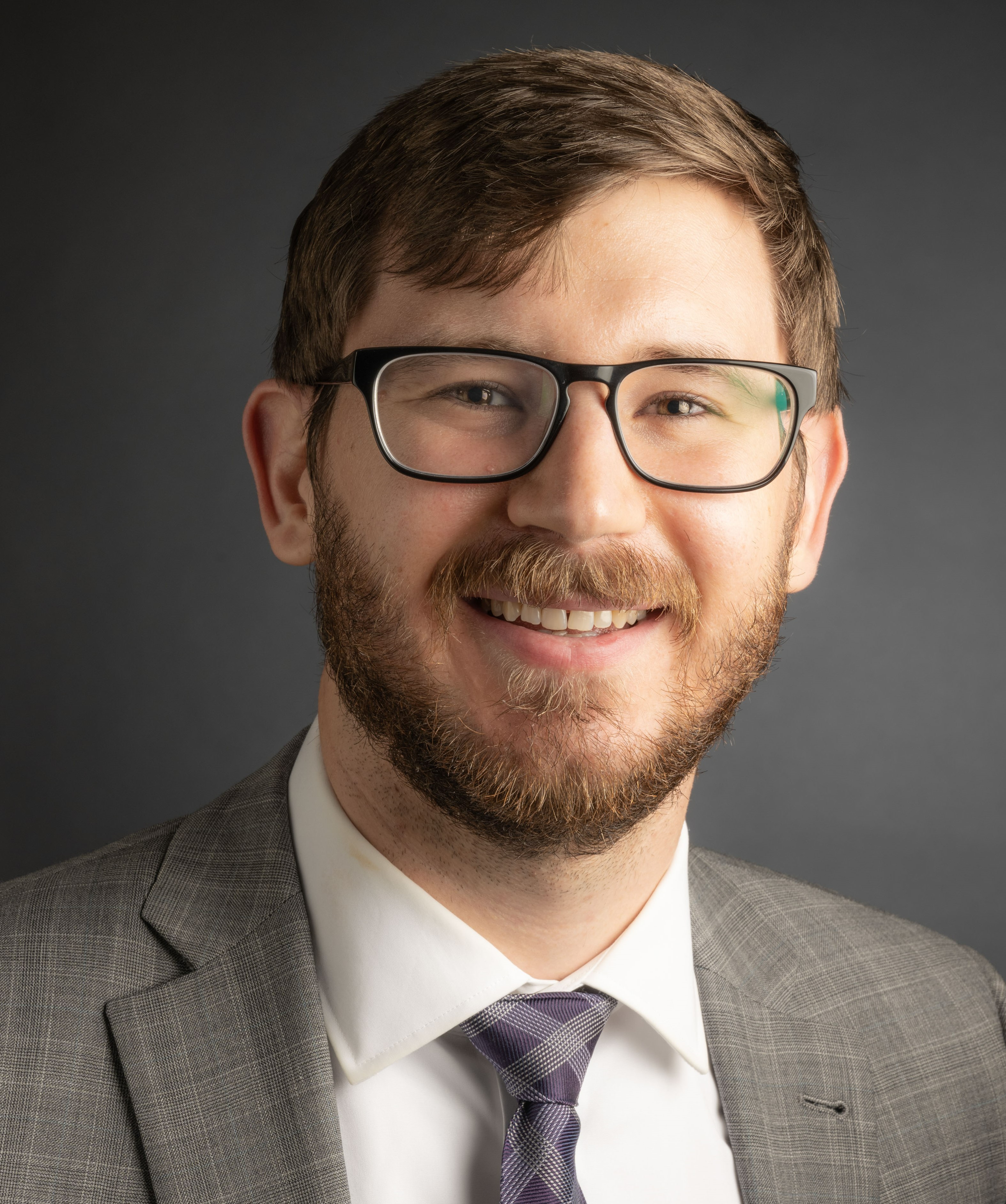}}]{Michael Zuzak} (Member, IEEE) received the Ph.D. degree in electrical engineering from the University of Maryland, College Park, MD, USA, in 2022. He is an Assistant Professor with the Department of Computer Engineering, Rochester Institute of Technology, Rochester, NY, USA. His current research interests include hardware security, computer architecture, and electronic design automation.
\end{IEEEbiography}

\begin{IEEEbiography}
[{\includegraphics[width=1in,height=1.25in]{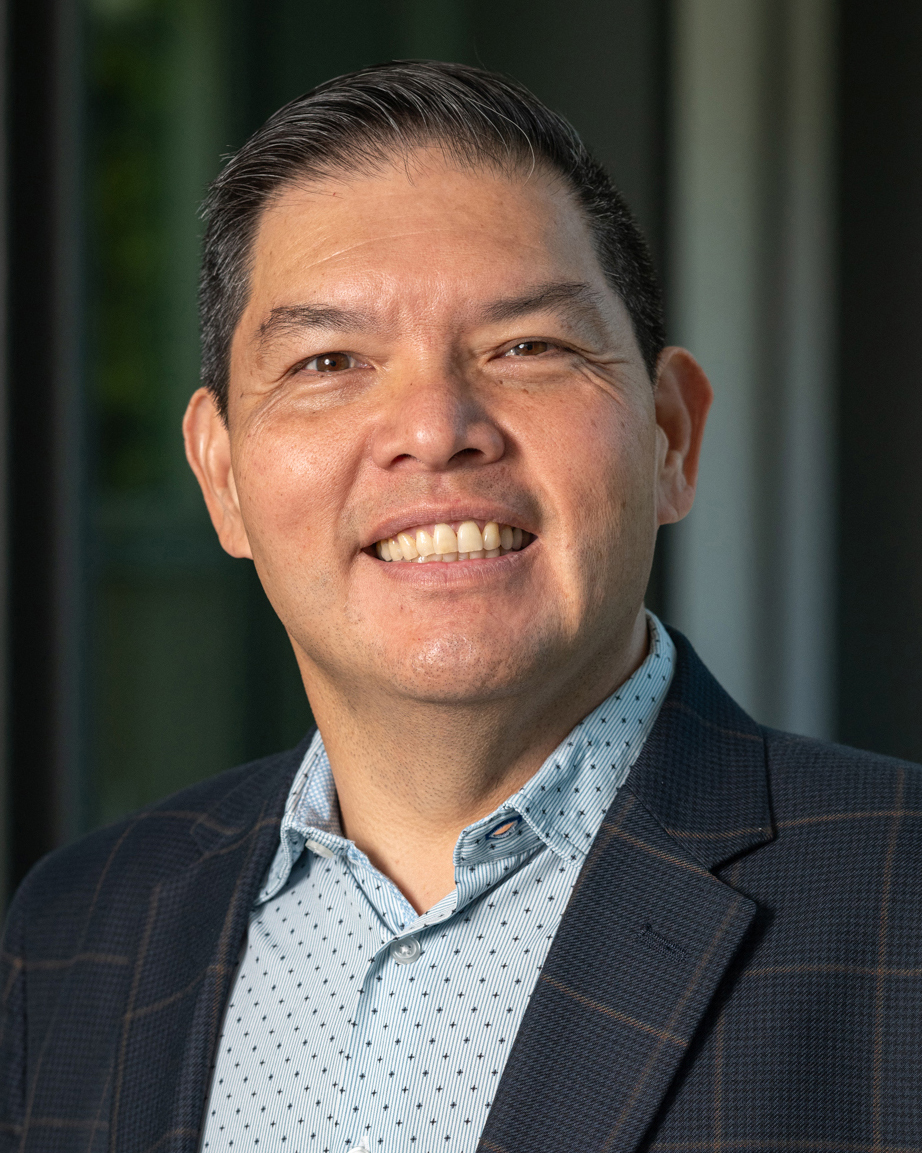}}]{Shanchieh Jay Yang} (Senior Member, IEEE) received his Ph.D. degree in electrical and computer engineering from the University of Texas, Austin, TX, USA, in 2001. He is the Reisenauer Family Director of the Institute for Informatics and Applied Technology at Gonzaga University. His research focuses on responsible artificial intelligence for cybersecurity and education. He was a 2019 NSF Trusted CI Open Science Fellow and a 2020 NSF Trusted CI TTP Fellow. He received the IEEE Region 1 Outstanding Teaching in an IEEE Area of Interest Award for outstanding leadership and contributions to cybersecurity and computer engineering in 2019. He was a co-chair for the IEEE Joint Communications and Aerospace Chapter in Rochester NY in 2005, when the chapter was recognized as an Outstanding Chapter of Region 1. 
\end{IEEEbiography}

\appendix

\section{Evaluation Module Attributes}
\label{appendix: eval_prompts}

The LLM must output three attributes: value, rationale, and provenance. `Value' must be one of: True Positive (TP) – correct prediction, False Positive (FP) – hallucination detected, or False Negative (FN) – failed retrieval based on the response and available evidence (NVD + CWE + hyperlinks). `Rationale' must explain why that value was chosen. `Provenance' must provide supporting evidence using document-chunk text from trusted sources (not LLM-generated summaries). Here are the engineered prompts for these attributes:

\vspace{2pt}

\begin{mybox}[frametitle=Evaluation Attributes]

\vspace{2pt}

\textit{\textbf{\small Value}}

\vspace{2pt}

{ \it \footnotesize
\noindent - Description: This attribute represents the accuracy of the response based on the provided context about the CVE-ID.

\noindent - Allowed Values: 

\noindent     - `TP' (True Positive): The response fully and accurately reflects the information in the context.
    
 \noindent    - `FP' (False Positive): The response contains inaccurate or incorrect information that are not supported by the context.
    
\noindent    - `FN' (False Negative): The response omits information that is present in the context.
    
\noindent - Requirements:

 \noindent    - Must strictly be one of the allowed values: `TP', `FP', or `FN'.
    
 \noindent    - The selection must be based on an objective comparison between the response and the context.
    
 \noindent    - Must follow the guidelines for Provenance.
}

\vspace{4pt}

\noindent \textit{\textbf{\small Rationale}}

{ \it \footnotesize
 \noindent            - Description: This attribute provides the reasoning behind the selected "value" for the evaluation of the response.
    
}

\vspace{4pt}

\noindent \textit{\textbf{\small Provenance}}

{ \it \footnotesize
 \noindent        - Detailed Instructions for Provenance based on the selected "value":

        \vspace{2pt}
        
  \noindent           - If "value" is `TP':
            
      \noindent           1. Carefully compare the response with the context.
                
      \noindent           2. Identify and extract the key segments from both the response and the context that align perfectly.
                
       \noindent          3. Format the provenance as follows:
                
          \noindent           - `response: "[Extracted segment from the response that matches the context]".'
                    
        \noindent             - `context: "[Corresponding segment(s) from the context]".'

            \vspace{2pt}
            
   \noindent          - If "value" is `FP':
            
    \noindent             1. Review the response thoroughly to identify parts that are incorrect or not present in the context.
                
    \noindent             2. Format the provenance as follows:
                
     \noindent                - `response: "[Incorrect or unsupported segment from the response]".'
                    
       \noindent              - `context: "[The closest matching part from the context, or indicate `No corresponding information in context' if none exists]".'

            \vspace{2pt}
            
   \noindent          - If "value" is `FN':
            
     \noindent            1. Examine the context to find essential information that is present but omitted in the response.
                
     \noindent            2. Format the provenance as follows:
                
      \noindent               - `response: "[The entire response]".'
                    
       \noindent              - `context: "[Relevant segment from the context that should have been matched]".'

            \vspace{2pt}
            
 \noindent            Guidelines:
            
    \noindent             - In all cases, ensure the provenance provided is clear, concise, and directly supports the selected "value".
                
     \noindent            - The provenance should be an exact match or a direct comparison between the response and the context, following the formatting rules strictly.

}

\end{mybox}

\section{Summarizing Technique Retrieval}
\label{appendix: relevancy}

The following shows the prompt used in {\textit Retr. LLM} to assess each URL's relevancy, and, if relevant, the LLM will summarize the URL content specifically to address either the  Exploitation or the Mitigation question.

\begin{mybox}[frametitle=Relevancy Prompts]

\noindent \textit{\textbf{\small Exploitation Prompt}}

{ \it \footnotesize 

\noindent You are a cybersecurity expert. Your task is to analyze the provided URL content for the [CVE-xxx-xxx] and provide a detailed summary.

        \noindent Content: {content}

        \noindent Please follow the steps below:

        \noindent Step 1: Assess Relevancy
        
   \noindent      - Does the content provide relevant information to describe how this [CVE-xxx-xxx] can be exploited?
        
  \noindent       - Answer: [Yes/No]

       \noindent  Step 2: Summarize Relevant Information
        
   \noindent      - If the answer is "Yes" in Step 1:
        
    \noindent         - Summarize the content with step-by-step description to exploit this vulnerability.
            
    \noindent     - If the answer is "No" in Step 1:
        
   \noindent          - Summary: NONE
}

\vspace{4pt}

\noindent \textit{\textbf{ \small Mitigation Prompt}}

{ \it \footnotesize 

\noindent You are a cybersecurity expert. Your task is to analyze the provided URL content for the [CVE-xxx-xxx] and provide a detailed summary.

        \noindent Content: {content}

        \noindent Please follow the steps below:

        \noindent Step 1: Assess Relevancy
        
   \noindent      - Does the content provide relevant information to describe the potential mitigation strategies for this vulnerability?
        
  \noindent       - Answer: [Yes/No]

        \noindent Step 2: Summarize Relevant Information
        
  \noindent       - If the answer is "Yes" in Step 1:
        
  \noindent           - Summarize the content to describe the potential mitigation strategies for this CVE.
            
  \noindent       - If the answer is "No" in Step 1:
        
  \noindent           - Summary: NONE
}

\end{mybox}

\EOD

\end{document}